%Paper: cond-mat/9510114
%From: Kiyomi Okamoto <kokamoto@stat.phys.titech.ac.jp>
%Date: Fri, 20 Oct 1995 15:12:42 +0900

%%%% This is a Plain TeX sourse file of
%%%% "DIMER CORRELARIONS IN SPIN-1/2 ALTERNATING XY CHAIN"
%%%% by Yohei Saika and Kiyomi Okamoto
%
%
\parindent=0.4cm
\font\smallcap = cmcsc10
\topskip 2.5truecm
\centerline{\bf DIMER CORRELARIONS IN SPIN-1/2 ALTERNATING XY CHAIN}
\bigskip
\centerline{ Y${\rm \hat o}$hei S{\smallcap aika} }
\centerline{ Department of Electrical Engineering, }
\centerline{ Wakayama National College of Technology, }
\centerline{ Nada-cho, Gobo-shi, Wakayama 644, Japan }
\medskip
\centerline{ and }
\medskip
\centerline{ Kiyomi O{\smallcap kamoto} }
\centerline{ Department of Physics, Tokyo Institute of
Technology, }
\centerline{ Oh-okayama, Meguro-ku, Tokyo 152, Japan }
\bigskip
\centerline{ ( Received ~~~~~~~~~~~~~~~~~~~~) }
\par
\bigskip
\bigskip
We investigate the long-distance asymptotic behavior of the dimer
correlations in the spin-$1/2$ alternating $XY$chain both at $T=0$
and at sufficiently low-temperatures.
The correlations consist of the dimer long-range order part and the
exponentially decaying one.
Although the dimer long-range order takes the different values
depending on the choice of the spin pairs, the behavior in the
decaying term is same irrespective of the choice of spin pairs.

\vfill\eject
\topskip0truecm
%%%%%%%%%%%%%%%%%%%%%%%%%%%%%%%%%%%%%%%%%%%%%%%%%%%%%%%%%%%%%%%%%%%%
The spin-Peierls phase transition has been studied as one of the
attracting problems by many theoretical and experimental phyisicists.
Some phyicists attempted to understand the mechanism of the
spin-Peierls phase transition by investigating the spin-$1/2$
alternating quantum Heisenberg chain.
This system is thought to represent the spin degree of freedom
of the organic compounds which fall into the spin-Peierls state
at sufficiently low-temperatures.
The degree of the bond alternations is treated as a given parameter,
though it should be determined so as to minimize the total energy.
\par
The spin-$1/2$ alternating $XY$ chain has been studied as one of the
exactly solvable models for the spin-Peierls state.
The Hamiltonian is
$$
H = J \sum_{j=1}^N \left[ 1 + (-1)^j \delta \right]
( S_j^x S_{j+1}^x + S_j^y S_{j+1}^y ),
\eqno(1)
$$
where $N$ is the system size which is assumed to be even and $\delta$
generates the bond alternation.
Here we only show the typical previous works on this model in the
following.
Pincus [1] first calculated the excitation spectrum by using the
Jordan-Wigner transformation.
He [1] pointed out that the finite energy gap is generated above
the ground state as far as $\delta > 0.$
Since then the phyisical quantities, such as the longitudinal spin
correlations $\langle S_i^z S_j^z \rangle$ and susceptibility,
have been evaluated by various methods.
The long-distance asymptotic behavior of the longitudinal
spin correlations was exactly obtained both at $T=0$ and at
sufficiently low-temperatures by one of the present authors
(K.O.).[2,3]
He [2,3] obtained the correlation length and the pre-exponential
factor exactly.
\par
The purpose of the present paper is to excatly estimate the
long-distance asymptotic behavior of the dimer correlations both
at $T=0$ and at sufficiently low-temperatures.
The definition of the dimer correlation is given as
$$
D(i:j) \equiv \langle T(i,i+1) T (j,j+1) \rangle, \eqno(2)
$$
where $T(l,m)$ is the dimer operator
$$
T(l,m) = S_l^+ S_m^- + S_l^- S_m^+.
\eqno(3)
$$
It is seen that the dimer correlations are classified into three kinds
depending on the choice of the spin pairs.
We point out the following statements:
(i)~The long-distance asymptotic behaviors of the dimer correlations
are given as the sum of the dimer long-range order part and the
exponentially decaying one;
(ii)~It takes different value of the dimer
long-range order depending on the choice of the spin pairs;
(iii)~It exhibits same behavior
both in the correlation length and the pre-exponential factor
irrespective of the choice of the spin pairs.
\par
The organization of the present paper is in the following.
We first show the diagonalization procedure of the spin-$1/2$ quantum
$XY$ chain with bond alternation.
We secondly show the long-distance asymptotic behavior of the
two-point correlation functions of fermion operators at $T=0$ and
at sufficiently low-temperatures.
Thirdly we estimate the long-distance asymptotic behavior of the dimer
correlations both at $T=0$ and at sufficiently low-temperatures.
\par
Let us begin with the diagonalization procedure.
On the first step, we rewrite the Hamiltonian (1) into the fermion
representation by the following Jordan-Wigner transformation
$$
\eqalignno{
& S_{2j-1}^+ = a_{2j-1}^{\dagger} K(2j-1) =
\left\{ S_{2j-1}^- \right\}^{\dagger}, \cr
& S_{2j-1}^z = a_{2j-1}^{\dagger} a_{2j-1} - {1\over 2}, & (4) \cr
& K(2j-1)
= \exp \left[ i\pi \sum_{l=1}^{j-1} a_{2l-1}^{\dagger} a_{2l-1}
              + i\pi \sum_{l=1}^j b_{2l}^{\dagger} b_{2l} \right], \cr
&& (5) \cr
& S_{2j-1}^+ = b_{2j}^{\dagger} K(2j)
= \left\{ S_{2j-1}^- \right\}^{\dagger},
S_{2j}^z = b_{2j}^{\dagger} b_{2j} - {1\over 2}, & (6) \cr
& K(2j)
= \exp \left[ i\pi \sum_{l=1}^j a_{2l-1}^{\dagger} a_{2l-1}
              + i\pi \sum_{l=1}^j b_{2l}^{\dagger} b_{2l} \right], \cr
&& (7)
}
$$
where $a_l$ and $b_m$ are the fermion operators and $K(l)$
is the kink operator.
By this transformation, the Hamiltonian (1) is rewritten into
$$
\eqalignno{
H & = J ( 1 - \delta ) \sum_{j=1}^{N/2}
  ( a_{2j-1}^{\dagger} b_{2j} - a_{2j-1} b_{2j}^{\dagger} ) \cr
  & + J ( 1 + \delta ) \sum_{j=1}^{N/2}
  ( a_{2j+1}^{\dagger} b_{2j} - a_{2j+1} b_{2j}^{\dagger} ).
& (8)
}
$$
On the next step, we transform the model Hamiltonian by use of the
canonical transformation after the Fourier transformation:
$$
\eqalignno{
& a_k = {1\over \sqrt{2}} ( \alpha_k + \beta_k ) e^{-i\theta_k},~~
  b_k = {1\over \sqrt{2}} ( \alpha_k - \beta_k ) e^{-i\theta_k}, \cr
&& (9) \cr
& \tan 2\theta_k = - \delta \tan k, & (10)
}
$$
where
$$
\eqalignno{
& a_k = {1\over N} \sum_k e^{ik(2j-1)} a_{2j-1},
  b_k = {1\over N} \sum_k e^{ik2j} b_{2j}. \cr
&&(11)
}
$$
Here the summation runs over the first Brillouin zone and $\alpha_k$
and $\beta_k$ are also the fermion operators.
By use of the above relations, the model Hamiltoninian is diagonalized
as
$$
\eqalignno{
& H = J \sum_k \omega_k \alpha_k^{\dagger} \alpha_k
    - J \sum_k \omega_k \beta_k^{\dagger} \beta_k, & (12) \cr
& \omega_k = \sqrt{ 1 - ( 1 - \delta^2 ) \sin^2 k }. & (13)
}
$$
We can see from (12) and (13) that the ground state is defined as the
half-filled state and that non-zero $\delta$ value generates the
energy gap above the ground state.
It also indicates that the Luttiger liquid state which is the ground
state of the spin-$1/2$ uniform $XY$ chain is changed into the
effective dimer state as far as $\delta > 0.$
The support for the above statements is, as Okamoto [2] calculated,
obtained by the long-distance asymptotic behavior of the longitudinal
spin correlation function $\langle S_i^z S_j^z \rangle.$
He showed that the correlation length becomes finite
when $\delta$ is non-zero value.
\par
Then we show the long-distance asymptotic behavior of the following
two-point correlation functions of fermion operators
$g(2m+1)$ which is given by
$$
\eqalignno{
& g(2m+1) \cr
\equiv & \langle a_{2j-1} b_{2j+2m}^{\dagger} \rangle
= - \langle a_{2j-1}^{\dagger} b_{2j+2m} \rangle \cr
= & {1\over 2} ( 1 + \delta ) L( 2m ) + {1\over 2} ( 1 - \delta )
L( 2m+2 ),
& (14)
}
$$
where
$$
\eqalignno{
  & L(2m)_{T=0} = {1\over 2N} \sum_k
{ \cos 2km \over \sqrt{ 1 - ( 1 - \delta^2 ) \sin^2 k } },
& (15) \cr
  & L(2m)_{T\not= 0} \cr
  & = {1\over 4N} \sum_k
  { \cos 2km \over \sqrt{ 1 - ( 1 - \delta^2 ) \sin^2 k } }
\tanh \left( {\omega_k \over 2\tilde T } \right). & (16)
}
$$
Here $\tilde T = T/J.$
The long-distance asymptotic behavior of $L(2m)_{T=0}$ and
$L(2m)_{T\not= 0}$ is exactly estimated by Okamoto.[2,3]
Here we only write down the asymptotic form of $L(2m)_{T=0}$
without entering into the details of the derivation.
$$
L(2m)_{T=0} \sim { (-1)^m\over 2 \sqrt{ \pi \delta } }
{ 1 \over \sqrt{m} } \exp \left[ - m a_{T=0} \right],
\eqno(17)
$$
where
$$
a_{T=0} = 1/\xi_{T=0} =
\log ( 1 + \delta ) - \log ( 1 - \delta ).
\eqno(18)
$$
Here $\xi_{T=0}$ is the correlation length.
When $\delta = 0,$ the correlation length becomes infinite, which
suggests the power-low behavior of $L(2m)_{T=0}.$
\par
We show the long-distance asymptotic behavior of $L(2m)_{T\not= 0}$
at sufficiently low temperatures:
$$
L(2m)_{T\not= 0} \sim
{
( -1 )^m 2 \tilde T u_1^m
\over
\sqrt{
\delta^2 + ( 1 + \delta^2 ) \pi^2 \tilde T^2 + \pi^4 \tilde T^4
}
},
\eqno(19)
$$
where
$$
\eqalignno{
u_1 = & { 1 + \delta^2 + 2 \pi^2 \tilde T^2
+ 2 \sqrt{
\delta^2 + ( 1 + \delta^2 ) \pi^2 \tilde T^2 + \pi^4 \tilde T^4 }
\over
1 - \delta^2
}.
\cr
&& (20)
}
$$
The above result gives the reasonable value at sufficiently low
temperatures, because the temperature dependence is evaluated
by the most contributing pole at sufficiently low temperatures.
If we expand the correlation length around $T=0,$ the derivation
of the correlation length at low temperatures can be estimated.
\par
The dimer correlations (Fig.1) are expressed by use of $g'$s as
follows;
$$
\eqalignno{
& D( 2j-1 : 2j-1+2m ) \cr
& \equiv \big\langle T( 2j-1, 2j ) T( 2j-1+2m, 2j+2m ) \big\rangle \cr
& = \big[ g(1) \big]^2 - g( 1 + 2m ) g( 1 - 2m ), & (21{\rm a}) \cr
& D( 2j-1 : 2j-1+2m ) \cr
& \equiv \big\langle T( 2j-1, 2j ) T( 2j-1+2m, 2j+2m ) \big\rangle \cr
& = \big[ g(-1) \big]^2 - g( -1 + 2m ) g( -1 - 2m ),
& (21{\rm b}) \cr
&  D( 2j-1 : 2j-1+2m ) \cr
& \equiv \big\langle T( 2j-1, 2j ) T( 2j-1+2m, 2j+2m ) \big\rangle \cr
& = g(1) g(-1) - g( 1 - 2m ) g( -1 + 2m ), & (21{\rm c})
}
$$
where
$$
T( l, m ) = S_l^+ S_m^- + S_l^- S_m^+.
\eqno(22)
$$
We first treat $T=0$ case.
The long-distance asymptotic behaviors of the dimer correlations are
obtained from eqs. (14), (17) and (18) through the straightforward
calculations:
$$
\eqalignno{
     & D( 2j-1: 2j-1+2m ) \cr
& \sim \left[ g(1) \right]^2
+ { 1 \over 8\pi } { 1 \over (2m)^2 }
\exp \left[ -2ma_{T=0} \right],
& (23{\rm a}) \cr
     & D( 2j: 2j+2m ) \cr
& \sim \left[ g(-1) \right]^2
+ { 1 \over 8\pi } { 1 \over (2m)^2 }
\exp \left[ -2ma_{T=0} \right],
& (23{\rm b}) \cr
     & D(2j-1: 2j+2m ) \cr
& \sim g(1) g(-1) + { 1 \over 8\pi } { 1 \over (2m+1)^2 }
\exp \left[ -(2m+1) a_{T=0} \right]. \cr
&~~~~&  (23{\rm c})
}
$$
The dimer correlations are sum of the dimer long-range order parts
and the exponentially decaying ones.
We should notice that the correlations have the same correlation
length and the same pre-exponential factor though they have different
values of the dimer long-range order.
The correlation length of the dimer correlations is exactly equal to
that of the longitudinal spin correlations derived by Okamoto.[2]
That is, the correlation length is given as
$$
\xi_{T=0} = 1/\left[ \log ( 1 + \delta ) - \log ( 1 - \delta )
\right].
\eqno(24)
$$
The correlation length grows from $0$ to $\infty$ with the decrease
in the parameter $\delta$ from $1$ to $0.$
At $\delta = 1$ where the system is the ensumble of the interacting
two-spin systems, it is natural that there is no correlation between
the different spin pairs.
In particular, both in $\delta$ $\rightarrow$ $1$ limit and in
$\delta$ $\rightarrow$ $0$ limit the correlation length behaves as
$$
\eqalignno{
\xi_{T=0}
& \rightarrow 1/\log [ 2 ( 1 - \delta ) ],~~( \delta \rightarrow 0 )
& (25{\rm a}) \cr
& \rightarrow 1/2\delta.~~~~~~~~~~~~~~~~
( \delta \rightarrow 1 )
& (25{\rm b})
}
$$
Finally at $\delta = 0$ where the system is reduced the spin-$1/2$
uniform $XY$ chain, the correlation length beoomes infinity.
\par
Next we discuss the dimer long-range order terms.
As we mentioned above, the dimer long-range order takes different
value by the choice of the spin pairs because it is constructed by the
product of the nearest neighbor correlations.
The explicit forms of dimer long-range order are represented by use of
the complete elliptic integrals of first and second kinds as follows.
$$
\eqalignno{
& \left[ g(1) \right]^2
= { 1 \over \pi^2 ( 1 + \delta )^2 }
\left[
\delta K( \sqrt{ 1 - \delta^2 } ) + E( \sqrt{ 1 - \delta^2 } )
\right], \cr
&~~& (26) \cr
& \left[ g(-1) \right]^2
= { 1 \over \pi^2 ( 1 - \delta )^2 }
\left[
- \delta K( \sqrt{ 1 - \delta^2 } ) + E( \sqrt{ 1 - \delta^2 } )
\right],\cr
&~~& (27)
}
$$
where
$$
\eqalignno{
  & K( \lambda )
= \int_0^{\pi} dk { 1 \over \sqrt{ 1 - \lambda^2 \sin^2 k } },
& (28) \cr
& E( \lambda ) = \int_0^{\pi} dk \sqrt{ 1 - \lambda^2 \sin^2 k }.
& (29)
}
$$
Here we examine the dimer long-range order both in
$\delta$ $\rightarrow$ $0$ limit and in $\delta$ $\rightarrow$ $1$
limit.
In $\delta$ $\rightarrow$ $0$ limit, we see
$$
\eqalignno{
& \lim_{m \rightarrow \infty} D( 2j-1 : 2j-1+2m )
= { 4 \over \pi^2 } \left( 1 + \delta \log \delta \right), \cr
&& (30{\rm a}) \cr
& \lim_{m \rightarrow \infty} D( 2j : 2j+2m )
= { 4 \over \pi^2 } \left( 1 - \delta \log \delta \right), \cr
&& (30{\rm b}) \cr
& \lim_{m \rightarrow \infty} D( 2j-1 : 2j+2m )
= { 4 \over \pi^2 } \left( 1 - ( \delta \log \delta )^2 /2 \right), \cr
&& (30{\rm c})
}
$$
and in $\delta \rightarrow 1$ limit
$$
\eqalignno{
& \lim_{m \rightarrow \infty} D( 2j-1 : 2j-1+2m )
= ( 1 - \delta )^2,
& (31{\rm a}) \cr
& \lim_{m \rightarrow \infty} D( 2j : 2j+2m )
= 4 - 2 ( 1 - \delta^2 ),
& (31{\rm b}) \cr
& \lim_{m \rightarrow \infty} D( 2j-1 : 2j+2m )
= 2 ( 1 - \delta ).
& (31{\rm c})
}
$$
\par
Then we discuss the temperature dependence of the dimer correlations
by using (19) and (20).
As we have stated above, (19) and (20) are suitable in the
sufficiently low temperature region.
The results are
$$
\eqalignno{
& D( 2j-1 : 2j-1+2m ) \cr
& \sim \left[ g(1)_{T\not= 0} \right]^2
+ { \tilde T^2 \over 8\pi }
\exp \left[ -2m/\xi_{T\not= 0} \right],
& (32{\rm a}) \cr
& D( 2j : 2j+2m ) \cr
& \sim \left[ g(-1)_{T\not= 0} \right]^2
+ { \tilde T^2 \over 8\pi }
\exp \left[ -2m/\xi_{T\not= 0} \right],
& (32{\rm b}) \cr
& D( 2j-1 : 2j+2m ) \cr
& \sim g(1)_{T\not= 0} g(-1)_{T\not= 0}
+ { \tilde T^2 \over 8\pi }
\exp \left[ -(2m+1)/ \xi_{T\not= 0} \right], \cr
&& (32{\rm c})
}
$$
where
$$
\xi_{T\not= 0} = \xi_{T=0}
\left( 1 - { \pi^2 \xi_{T=0} \tilde T^2 \over \delta } \right).
\eqno(33)
$$
Here $\xi_{T\not= 0}$ denotes the correlation length at sufficiently
low temperatures.
We can see that the $\xi_{T\not= 0} < \xi_{T=0}$ due to the thermal
effects.
It means that there is no second-order phase transition at finite
temperatures, as the Mermin-Wagner theorem guarantees.
The correlations at sufficiently low temperatures have similar
character in comparison with those at $T=0$.
That is, the dimer long-range order takes different value depending on
the choice of spin pairs, but the exponential decaying part has the
same behavior in the correlation length and pre-exponential factor
in each correlation.
\par
In summary, we obtain the following statements:
(i)~The dimer long-range order takes the different values depending on
the choice of spin pairs;
(ii)~The behavior in the decaying term is same irrespective of the
choice of spin pairs, both at $T=0$ and at sufficiently low
temperatures.
%%%%%%%%%%%%%%%%%%%%%%%%%%%%%%%%%%%%%%%%%%%%%%%%%%%%%%%%%%%%%%%%%%%%%
\vskip 3.00truept
%%%%%%%%%%%%%%%%%%%%%%%%%%%%
\bigskip
\bigskip
\leftline{ References }
%%%%%%%%%%%%%%%%%%%%%%%%%%%%
\vskip 16.0truept
%%%%%%%%%%%%%%%%%
\par
1.~P.~Pincus,~Solid~State~Commun.~{\bf 9,}~71~(1971).
\par
2.~K.~Okamoto,~J.~Phys.~Soc.~Jpn.~{\bf 57,}~2947~(1988).
\par
3.~K.~Okamoto,~J.~Phys.~Soc.~Jpn.~{\bf 59,}~4286,(1990).
%%%%%%%%%%%%%%%%%%%%%%%%%%%%%%%%%%%%%%%%%%%%%%%%%%%%%%%%
\vfill\eject
%%%%%%%%%%%%%%%%%%%%%%%%%%%
\leftline{ Figure Caption }
%%%%%%%%%%%%%%%%%%%%%%%%%%%
\vskip 16.0truept
%%%%%%%%%%%%%%%%%
{\parindent 0.0truecm
Fig.1~~~~~~Three kinds of the dimer correlations.
(a),(b) and (c) correspond to eq.(21a),(21b) and (21c), respectively.

\end